\documentclass[sigconf,nonacm]{acmart}
\AtBeginDocument{%
  }

\setcopyright{none}
\renewcommand\footnotetextcopyrightpermission[1]{}

\usepackage{siunitx}
\usepackage{multirow}
\usepackage{booktabs}
\usepackage{graphicx}
\usepackage{epstopdf}		
\usepackage{textcomp}

\usepackage{standalone}

\usepackage{tikz}
\usetikzlibrary{bayesnet}
\usetikzlibrary{external}

\usepackage{pgfplots}
\pgfplotsset{compat=newest}
\usetikzlibrary{bayesnet}
\usetikzlibrary{patterns}
\usetikzlibrary{pgfplots.groupplots}

\usetikzlibrary{chains,fit,shapes,positioning,arrows,arrows.meta}
\pgfdeclarelayer{background}
\pgfsetlayers{background,main}

\pgfplotsset{
  select row/.style={
    x filter/.code={\ifnum\coordindex=#1\else\fi}
  }
}

\pgfplotscreateplotcyclelist{six-bars-list}{
{six-bars-col1,fill=six-bars-col1},
{six-bars-col2,fill=six-bars-col2},
{six-bars-col3,fill=six-bars-col3},
{six-bars-col4,fill=six-bars-col4},
{six-bars-col5,fill=six-bars-col5},
{six-bars-col6,fill=six-bars-col6}
}

\usetikzlibrary{decorations.markings}
\usetikzlibrary{decorations}
\usetikzlibrary{decorations.pathreplacing,calc}

\tikzstyle{request}=[draw=none, rectangle, fill=red!20!green!10!gray!20,align=center,left]
\tikzstyle{light}=[draw=none, rectangle, fill=blue!30!gray!10,align=center]
\tikzstyle{dark}=[draw=none, rectangle, fill=blue!10!gray!40,align=center]

\usepackage[T1]{fontenc}
\usepackage[utf8]{inputenc}

\usetikzlibrary{matrix}
\usetikzlibrary{arrows}
\usepackage{filecontents}

\usepackage{xspace}
\newcommand{\etal}{{et al.\@\xspace}}

\usepackage[olditem]{paralist}

\usepackage{wrapfig}
\usepackage{url}
\usepackage{subfig}

\usepackage{float}
\newfloat{copyrightbox}{h}{junk}

\usepackage[inline]{enumitem}
\setlist{nosep, topsep=0pt, leftmargin=2em}

\usepackage{amsmath}

\newcommand{\fsc}{\textsc}

\newcommand{\fsl}{\textsl}

\begin{document}

\title{PCap: Personalized Retrieval-Stage Diversity Capping in Facebook Marketplace}

\author{Guangchao Yuan}
\affiliation{%
  \institution{Meta}
  \city{Menlo Park}
  \state{CA}
  \country{USA}
}
\email{gcyuan@meta.com}

\author{Janis Fuh}
\affiliation{%
  \institution{Meta}
  \city{Menlo Park}
  \state{CA}
  \country{USA}
}
\email{jfuh@meta.com}

\author{Christopher Choate}
\affiliation{%
  \institution{Meta}
  \city{Boston}
  \state{MA}
  \country{USA}
}
\email{cchoate@meta.com}

\author{Xun Tang}
\affiliation{%
  \institution{Meta}
  \city{Menlo Park}
  \state{CA}
  \country{USA}
}
\email{xuntang@meta.com}

\author{Wenqi Zhu}
\affiliation{%
  \institution{Meta}
  \city{Menlo Park}
  \state{CA}
  \country{USA}
}
\email{wqzhu@meta.com}

\author{Chengyi Zhang}
\affiliation{%
  \institution{Meta}
  \city{Menlo Park}
  \state{CA}
  \country{USA}
}
\email{zhangcy@meta.com}

\author{Pavan Kumar Paalya Chandrashekar}
\affiliation{%
  \institution{Meta}
  \city{Menlo Park}
  \state{CA}
  \country{USA}
}
\email{pavanpc@meta.com}

\author{Jiang Han}
\affiliation{%
  \institution{Meta}
  \city{Menlo Park}
  \state{CA}
  \country{USA}
}
\email{jianghan1@meta.com}

\author{Jiangyuan Li}
\affiliation{%
  \institution{Meta}
  \city{Menlo Park}
  \state{CA}
  \country{USA}
}
\email{jiangyuanli@meta.com}

\author{Hongyan Wang}
\affiliation{%
  \institution{Meta}
  \city{Menlo Park}
  \state{CA}
  \country{USA}
}
\email{hongyan@meta.com}

\author{Shuting Wang}
\affiliation{%
  \institution{Meta}
  \city{Menlo Park}
  \state{CA}
  \country{USA}
}
\email{shutingwang@meta.com}

\renewcommand{\shortauthors}{Yuan et al.}

\begin{abstract}
  We propose a personalized capping framework (PCap) to improve the diversity in Facebook Marketplace by introducing user-level diversity constraints at the retrieval stage. PCap models individual diversity preferences using Shannon entropy-based scoring, segments users into diversity buckets, and applies personalized category caps during multi-source candidate retrieval. To navigate the high-dimensional parameter space of per-bucket caps, we leverage an automated online optimization method called Parameter Tuning Sequence. Large-scale online experiments demonstrate that PCap significantly improves users' browsing experience shown in engagement metrics.  This work provides practical insights into integrating personalized diversity into industrial retrieval systems.
\end{abstract}




\maketitle

\section{Introduction}
On Facebook Marketplace browse feed, users are recommended with relevant products that match their interests. The quality of recommendation directly impacts users' satisfaction and retention. However, the current recommender relies on users' past engagements, which could introduce a feedback loop on feeding top interested content and reduce the diversity of the recommendation. Diversity has been introduced alongside accuracy to capture the mutual influence between items, thereby improving the user satisfaction with recommendation lists \cite{Zhang+08,Ziegler+05}.

Several methods demonstrated the effectiveness of modeling the personalized diversity with the offline experiment on some public datasets \cite{Eskandanian+17,Di-Noia+14}. Wang {\etal} \cite{Wang+20} proposed an industry solution of applying the personalized diversity in the \emph{ranking stage} by adjusting the ranking item lists.  Li {\etal} \cite{Li+24} deployed diversification across full pipeline stages in a short-video platform, concluding that ranking is the most suitable stage. Our work focuses on retrieval-stage diversification and shows that user-level diversity preference can make retrieval-stage interventions effective in scalable recommendation systems. 



While personalized diversity is well-explored in ranking and offline settings, deploying it at the retrieval stage under strict latency and scalability constraints remains an open industry challenge. To bridge this gap, we present PCap, a \emph{Personalized Capping Framework} that improves diversity in Facebook Marketplace via the \emph{retrieval stage}, enabling more diverse content candidates into the ranking stage. PCap models user-level diversity preferences and applies personalized caps on candidates retrieved from each content group, enabling a more diverse candidate pool to reach downstream ranking models while respecting latency and scalability requirements.


Extensive experiments on real-world A/B tests on Facebook Marketplace population show that PCap improves users' browsing experience reflected in statistically significantly increased engagement metrics. Our main contributions are threefold:
\begin{enumerate}
\item \textbf{A scalable, retrieval-stage diversification framework}: we design and deploy a distributed capping mechanism that applies diversity constraints directly at the indexer-shard and aggregation levels. This ensures a balanced candidate pool reaches downstream ranking models while strictly preserving multi-source recall and low-latency requirements.
\item \textbf{Automated high-dimensional parameter tuning}: we 
\\demonstrate how to adapt a complex, high dimensional personalization problem---spanning user diversity weights, category buckets, and source multiplier---onto an automated Parameter Tuning Sequence (PTS) framework. This operationalizes continuous and grid-search-like optimization in live production, removing the bottleneck of manual parameter tuning.
\item \textbf{Empirical insights on "edge-heavy" personalization}: we validate PCap through large-scale online experiments, demonstrating statistically significant engagement improvements and revealing that personalized diversity primarily benefits users at preference extremes.
\end{enumerate}

\section{Methodology}
We design PCap to improve diversity by intervening during candidate retrieval within a multi-source framework. Specifically, PCap introduces personalized, group-level constraints that shape the candidate pool before ranking, enabling end-to-end diversity improvements while preserving relevance.


We leverage a manually-designed commerce taxonomy: \emph{Facebook Product Taxonomy (FPT)} to define groups. Every product is assigned with one of hundreds of FPT categories, e.g., ``Cell Phones \& Accessorie'', ``Cars \& Trucks'', ``Furniture''. The category of each product is predicted by an image-text classification model using annotated labels.  While we use FPT in this work, PCap supports any sparse, content-level features as grouping keys.

We first present our \fsl{personalized diversity modeling} approach and then detail how these signals are operationalized through two retrieval-stage capping mechanisms.

\subsection{Personalized Diversity Modeling}\label{sec:modeling}
Our approach is built on top of the hypothesis that users have different diversity preferences \cite{Wang+20,Eskandanian+17,Chen+13} when browsing products in Marketplace: users with limited interests may expect more similar products from a certain category; whereas users with diverse preferences may expect more diverse products across different categories.

Let $p(c|u)$ represent a user's clicked probability in FPT category c, $C$ denote all the categories, and $N$ denote the user's total number of clicked products ($N>1$). The original diversity metric in Marketplace is defined based on the normalized Herfindahl–Hirschman Index (HHI) \cite{HH-Index+25}. It measures the competition among different FPT shares $\frac{1-\sum_{c \in C} p(c|u)^2}{1 - \frac{1}{N}}$. However, this metric doesn't consider users' engagement levels. For example, user A clicks four products, each belonging to different FPT categories; user B clicks ten products, each belonging to different FPT categories. The diversity in terms of HHI for both user A and user B is 1.  We argue that user B should have a higher diversity score than user A since user B engages with more products.

We exploit Shannon entropy \cite{Wang+20,Eskandanian+17,Di-Noia+14} to overcome the limitation of HHI and integrate users' engagement into a user’s diversity score. Following \cite{Wang+20}, our personalized diversity modeling approach has the following three steps:
\begin{enumerate}
\item Given a user's recent click behavior over a sliding window on Marketplace feed, we obtain the Shannon entropy of each user u. A larger Shannon entropy indicates that a user historically engages with a broader variety of categories, which we use as a proxy for their diversity preference: 
\begin{equation}
S(u) = - \sum_{c \in C}{p(c|u) \times \log{p(c|u)}}
\end{equation}
\item We normalize the diversity scores of all users: 
\begin{equation}
d_u = \frac{S(U)-S(min)+l}{S(max)-S(min)+l}(l\geq 0)
\end{equation}
$S(max)$ means the max entropy value among all users, while $S(min)$ represents the min entropy value among all users. $l$ is a parameter to control the personalization degree. A larger $l$ indicates less personalization. We set $l$ as zero in the initial step to understand the impact of fully personalization before introducing additional smoothing. After normalization, all users' diversity scores would be between zero and one. 
\item We remove users with fewer clicks than a predefined threshold and divide the remaining users into six buckets based on their diversity scores: users in \fsc{bucket-1} are the least diverse, whereas users in \fsc{bucket-6} are the most diverse. There are two objectives when doing the bucketization: 
    \begin{enumerate*}
    \item users are roughly evenly distributed across six buckets;
    \item users within a bucket could mostly remain stable: an unstable user means her bucket would shift frequently day by day.
    \end{enumerate*}
We use these user-level diversity buckets to control how aggressively diversity constraints are applied during candidate retrieval, as described next. We adopt a fixed and small number of diversity buckets to balance personalization with system stability. In large-scale production systems, fixed buckets provide predictable behavior, low computational overhead, and robustness to noise in short-term user activity. 
\end{enumerate}

\subsection{Capping Mechanism}
In large-scale retrieval systems, candidate generation is distributed across multiple sources and shards to meet strict latency constraints. To maintain recall, each source independently over-fetches candidates, which can bias the aggregated candidate set toward dominant categories and reduce diversity.

PCap addresses this issue by introducing diversity-aware capping directly within the retrieval pipeline. Each retrieval source is subject to a fixed fetch cap. Candidates are retrieved from multiple indexer shards and aggregated, with over-fetching applied to maintain recall. Capping mechanisms are applied at both the shard-level \fsl{scanning} stage and the \fsl{aggregation} stage to reduce candidates in a controlled and diversity-aware manner.

\begin{figure*}[htbp]
    \centering
    \includegraphics[width=0.8\textwidth]{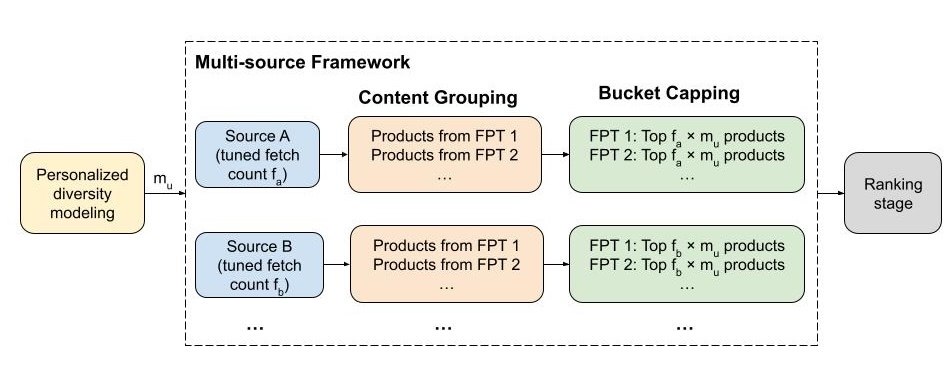}
    \caption{PCap framework at the Facebook Marketplace retrieval stage.}
    \label{fig:pcap}
\end{figure*}

\subsubsection{Content Grouping}
Candidates are grouped based on content-level features. In this work, we use FPT as grouping keys, though PCap supports any sparse content features. Grouping enables the system to reason about diversity explicitly and apply constraints at the group level rather than at the individual item level.

\subsubsection{Bucket Capping}
A bucket-level cap is applied to limit the number of candidates from each group. 

In PCap (Figure~\ref{fig:pcap}), the cap is determined by two factors: 
\begin{enumerate*}[label=(\arabic*)]
    \item the retrieval source fetch count scaled by a tuned multiplier $f_k (k \in K)$, where $K$ represents all the retrieval sources;
    \item a personalized diversity multiplier $m_u$ based on the user's diversity bucket (Section \ref{sec:modeling}). Users with lower diversity preference are assigned relaxed caps, while users with higher diversity preference receive tighter caps to promote broader category exposure.
\end{enumerate*}
These caps are applied consistently across retrieval stages to prevent any single group from dominating the candidate set while preserving high-quality and relevant items. The details of obtaining $m_u$ are described in Section \ref{sec:pts}. 


\section{Empirical Evaluations}
\subsection{Experimental Setup}
We evaluate PCap through two-phase large-scale online A/B tests on Facebook Marketplace. Each phase covers a substantial fraction of users over multiple weeks to ensure statistical reliability.  

We intentionally evaluate PCap exclusively through online A/B testing rather than traditional offline baselines. Because retrieval-stage capping fundamentally alters the candidate pool and shifts downstream ranking dynamics, offline metrics often fail to capture the true, second-order effects on live user behavior. Consequently, we rely entirely on large-scale online experiments to accurately measure how these retrieval interventions translate into actual user engagement and system-wide diversity.  

Additionally, established diversity methods such as Maximal Marginal Relevance (MMR) \cite{Carbonell+98} and determinantal Point Process (DPP) \cite{Chen+17} are designed for ranking-stage reranking over a fixed candidate set with full relevance scores available. They are systemically incompatible with retrieval-stage constraints, where candidates are distributed across multiple indexer shards under strict latency budgets and no global relevance scores exist for trade-off computation. We therefore compare against the most natural retrieval-stage baselines: no capping and uniform capping (a non-personalized variant of PCap):
\begin{itemize}
    \item \emph{No capping (control)}: the retrieval system without any diversity-aware capping mechanism. Candidates are retrieved purely based on source fetch counts.
    \item \emph{Uniform capping}: a simplified capping variant that segments users by buyer and non-buyer status, applying more aggressive caps for non-buyers and relaxed caps for buyers. The same capping rule is applied uniformly within each segment.
    \item \emph{PCap}: the full framework with six entropy-based diversity buckets and per-bucket multipliers optimized via PTS (Section \ref{sec:pts}).
\end{itemize}
    
The two-phase design proceeds as follows:
\begin{enumerate*}[label=(\arabic*)]
    \item \fsl{Phase 1} (uniform capping vs. no capping): validates whether retrieval-stage capping improves content exposure and engagement.
    \item \fsl{Phase 2} (PCap vs. uniform capping): measures the incremental value of personalized capping over uniform capping.
\end{enumerate*}

This design answers two key questions. First, does retrieval-stage capping improve the user experience? Second, does fine-grained personalization provide additional value beyond uniform capping?

\subsubsection{Evaluation Metrics}
We evaluate PCap using engagement, diversity, and system performance metrics. Results are reported as percentage change with $95\%$ confidence intervals.
\begin{itemize}
    \item Viewport Views (VPV): number of product listing views on Marketplace Feed per day. An increase indicates a positive browsing experience where users are willing to spend more time exploring the feed.
    \item Product Detail Page Clicks (PDP): number of user clicks into a listing's product detail page per day, reflecting active purchase consideration.
    \item Message Listing Interactions (MLI): number of buyer-seller conversations per day in which the seller replies, reflecting downstream purchase intent and transaction initiation.
    \item Marketplace Sessions: number of Facebook-linked sessions containing at least one Marketplace session per day, reflecting user return behavior.
    \item Diversity Score (HHI): higher scores indicate broader category exposure \cite{HH-Index+25}. We use Shannon entropy to model user preferences (as it captures engagement volume) and HHI to evaluate output diversity (as it measures category concentration and is natively supported in our production QE infrastructure).
    \item PF Latency: average latency of Marketplace Feed pagination requests,  measuring retrieval-stage overhead.
\end{itemize}

\subsection{Phase 1: Uniform Capping vs. No Capping}
To validate the premise that retrieval-stage diversity capping benefit users, we first test uniform capping against the no-capping production baseline. Uniform capping significantly increases VPV ($+0.3088\%$), confirming that category-level caps at the retrieval stage successfully expose users to more diverse content. However, deeper engagement metrics (PDP and MLI) and sessions show no statistically significant improvement. This gap between exposure and engagement suggests that diversity applied uniformly, without regard to individual user preferences, fails to convert increased content variety into meaningful user actions. This observation directly motivates the personalized approach in PCap.

\begin{table}
  \caption{Online A/B testing results of uniform capping}
  \label{tab:phase-1}
  \begin{tabular}{ll}
    \toprule
    Metrics & Change\%\\
    \midrule
    Viewport Views & $+0.3088\%^{**}$ \\
    Product Detail Page Clicks & $+0.0313\%$ \\
    Message Listing Interactions & $-0.2604\%$ \\
    Marketplace Sessions & $+0.0469\%$ \\
    PF Latency  & $+8.6ms$ \\
  \bottomrule
  \multicolumn{2}{l}{\footnotesize **: statistically significant at 95\% confidence} \\[-.5ex]
\end{tabular}
\end{table}

\subsection{Parameter Tuning Sequence}\label{sec:pts}
A core operational challenge is determining the optimal multiplier ($m_u$) for each diversity bucket. While the ordinal ranking of expected diversity is known a priori from the bucket construction, the high-dimensional parameter space-spanning category bucket sizes, source multipliers, and user diversity weights---makes manual tuning infeasible at scale. Furthermore, as noted in our evaluation setup, offline replay consistently underestimates the second-order effects of capping on live session composition and downstream ranking dynamics, making online automated tuning a necessity.  

To resolve this and scale the system, we leveraged an existing automated online tuning framework known as the \emph{Parameter Tuning Sequence (PTS)}. By mapping our diversity parameters to this framework, we iteratively learned the optimal magnitude of bucket multipliers directly from live user behavior. Since uniform capping was implemented after \fsl{Phase 1}, it now serves as the baseline against which PCap is evaluated.

\subsubsection{Adapting PTS for Diversity Tuning}
\begin{figure}[h]
    \centering
    \includegraphics[width=0.48\textwidth]{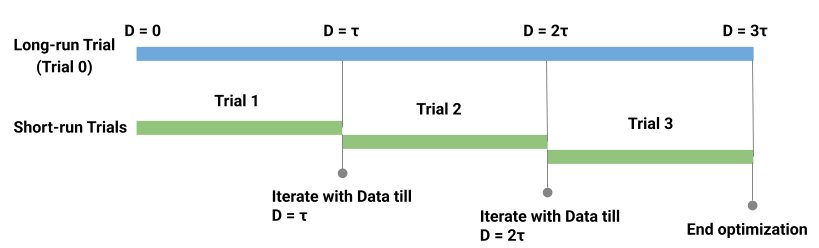}
    \caption{PTS framework.}
    \label{fig:pts}
\end{figure}

PTS operates as a sequential, range-narrowing A/B testing procedure, conceptually functioning as an online grid search (Figure \ref{fig:pts}).

Each trial arm corresponds to a distinct combination of bucket multipliers drawn from a pre-specified range. Each round holds a control arm fixed while exposing candidate arms to a small slice of traffic for several days ($\tau$). This duration is carefully calibrated to span at least one full weekday/weekend cycle, capturing the dominant source of variance in our target engagement metrics. After each round, the parameter envelope contracts toward the configurations that show the strongest performance, allowing the next round to sample more densely in that high-performing neighborhood.

\subsubsection{Production-Driven Design Choices}
We deliberately opted to utilize this grid-search-based infrastructure over implementing more complex, sample-efficient methods (e.g., Bayesian optimization or bandit-style allocation) for three practical industry reasons:
\begin{enumerate*}[label=(\arabic*)]
    \item The per-arm noise inherent in real-world engagement metrics is substantial; the computational overhead of complex acquisition machinery rarely pays off within our compute budget.
    \item A uniform grid produces transparent, monotonic results.
    \item The grid structure integrates seamlessly with our existing quantitative evaluation (QE) infrastructure, which is built around fixed-allocation arms rather than adaptive policies.
\end{enumerate*}

\subsubsection{Empirical Insights}
The automated tuning revealed a clear, monotonic pattern across the diversity spectrum, validating both our bucket construction and the underlying personalization hypothesis: the cap is a meaningful lever, and its optimal setting is genuinely a function of the user's intrinsic diversity preference.
\begin{enumerate*}[label=(\arabic*)]
    \item Users in the middle buckets (\fsc{bucket-3} and \fsc{bucket-4}) consistently converge at or near the neutral multiplier, indicating that additional personalization here yields no measurable engagement lift.
    \item Less diverse cohorts (\fsc{bucket-1} and \fsc{bucket-2}) require a larger multiplier---a relaxed cap that lets the system admit more items from any single FPT category.
    \item Higher diverse cohorts (\fsc{bucket-5} and \fsc{bucket-6}) converge on a smaller multiplier, tightening the cap and successfully forcing broader category exploration across the inventory.
\end{enumerate*}


\subsection{Phase 2: PCap vs. Uniform Capping}
Using the PTS-optimized per-bucket multipliers, we evaluate PCap against uniform capping. The results of the online test (Table~\ref{tab:phase-2}) show consistent positive movement across all primary engagement metrics, with statistically significant improvements in VPV, PDP, and Marketplace Sessions. Compared with uniform capping, PCap produces larger gains on deeper engagement metrics, suggesting that personalized caps improve the alignment between retrieval-stage diversification and user intent.  PCap introduces only $0.8 \text{ms}$ additional latency over uniform capping, indicating that personalized diversity modeling can be deployed with minimal incremental compute overhead. The combined latency overhead of uniform capping and PCap remains within production latency requirements.

\begin{table}
  \caption{Online A/B testing results of PCap}
  \label{tab:phase-2}
  \begin{tabular}{ll}
    \toprule
    Metrics & Change\% \\
    \midrule
    Viewport Views & $+0.2243\%^{**}$ \\
    Product Detail Page Clicks & $+0.2250\%^{**}$ \\
    Message Listing Interactions & $+0.1109\%$ \\
    Marketplace Sessions & $+0.1708\%^{**}$ \\
    PF Latency  & $+0.8ms$ \\
  \bottomrule
  \multicolumn{2}{l}{\footnotesize **: statistically significant at 95\% confidence} \\[-.5ex]
\end{tabular}
\end{table}

\subsubsection{Diversity Analysis}
Table \ref{tab:diversity} shows the diversity score changes in the pagination feed (page 2 and later). The results suggest that PCap modulates diversity in the intended direction across the full spectrum: less diverse cohorts (\fsc{bucket-1} and \fsc{bucket-2}) receive more concentrated feeds aligned with their focused preferences, while higher diverse users (\fsc{bucket-5} and \fsc{bucket-6}) receive broader category coverage --- with \fsc{bucket-6} showing a significant $0.25\%$ increase. Medium-diversity users show minimal change in diversity scores, consistent with the PTS experiments.

\begin{table}
  \caption{Pagination feed diversity score by bucket}
  \label{tab:diversity}
  \begin{tabular}{ll}
    \toprule
    Bucket & Diversity Score Change\%\\
    \midrule
    \fsc{bucket-1} & $-1.0537\%^{**}$\\
    \fsc{bucket-2} & $-0.9195\%^{**}$\\
    \fsc{bucket-3} & $-0.2947\%^{**}$\\
    \fsc{bucket-4} & $-0.1116\%$\\
    \fsc{bucket-5} & $+0.0037\%$\\
    \fsc{bucket-6} & $+0.2469\%^{**}$\\
  \bottomrule
  \multicolumn{2}{l}{\footnotesize **: statistically significant at 95\% confidence} \\[-.5ex]
\end{tabular}
\end{table}

\subsection{Summary of Findings}
The two-phase evaluation yields three insights for exploiting personalized diversity in scalable retrieval systems.

First, exposure alone does not drive engagement. Uniform capping successfully increases the variety of content shown to users (\fsl{Phase 1}), but without matching diversity to individual preferences, this increased exposure does not translate into deeper engagement such as clicks or sessions. 

Second, personalization is effective at the preference extremes. From a production standpoint, navigating this parameter space demonstrated that personalization efforts should be strictly "edge-heavy". Rather than attempting to fine-tune the majority of users who represent the "global average", computational resources and architectural complexity are best justified by focusing on users at the extremes (high-focus or high-exploration).

Third, automated parameter discovery is useful at scale. The high-dimensional parameter space spanning multiple diversity buckets and retrieval sources makes manual tuning infeasible. Automating parameter discovery through a system like PTS is a fundamental prerequisite for deploying diversity at scale.

\section{Conclusion and Future Work}
We presented a personalized capping framework that improves diversity in the Facebook Marketplace by combining \fsl{personalized diversity modeling} with retrieval-stage capping mechanisms: \fsl{content grouping} and \fsl{bucket capping}. A two-phase evaluation on large-scale online experiments demonstrates that PCap significantly improves users' browsing experience shown in engagement metrics.

In future, we plan to explore the following areas:
\begin{itemize}
    \item Exploring other content grouping methods (e.g., co-engagement based grouping) to capture users' intuition beyond semantic categories.
    \item Leveraging the clustering approach to dynamically obtain the user diversity buckets \cite{Eskandanian+17}.
    \item Dynamically adjusting the capping thresholds of different categories to account for variations in inventory distribution.
    \item Applying a similar user model to cold-start users who lack sufficient engagement data to derive a personalized diversity score.
\end{itemize}


\bibliographystyle{ACM-Reference-Format}
\bibliography{diversity}

\appendix

\end{document}